%% file: sample-sigconf.tex
\documentclass[sigconf, anonymous=false]{acmart}

\AtBeginDocument{%
  }

\setcopyright{acmcopyright}
\copyrightyear{2023}
\acmYear{2023}
\acmDOI{XXXXXXX.XXXXXXX}

\acmConference[WWW '23]{International Conference Companion on World Wide Web}{April 30--May 4, 2023}{Austin, TX}
\acmPrice{15.00}
\acmISBN{978-1-4503-XXXX-X/18/06}

\newcommand{\rev}[1]{#1}

\usepackage{graphicx}  % DO NOT CHANGE THIS
\usepackage{balance}
\begin{document}

\title{Detecting Social Media Manipulation in Low-Resource Languages}


\author{Samar Haider, Luca Luceri, Ashok Deb, Adam Badawy, Nanyun Peng, Emilio Ferrara\\University of Southern California, Information Sciences Institute}

\renewcommand{\shortauthors}{Heider et al.}

\begin{abstract}
Social media have been deliberately used for malicious purposes, including political manipulation and disinformation. Most research focuses on  high-resource languages. However, malicious actors share content across countries and languages, including low-resource ones. 
Here, we investigate whether and to what extent malicious actors can be detected in low-resource language settings. We discovered that a high number of accounts posting in Tagalog were suspended as part of Twitter's crackdown on interference operations after the 2016 US Presidential election. 
By combining text embedding and transfer learning, our framework can detect, with promising accuracy, malicious users posting in Tagalog  without any prior knowledge or training on malicious content in that language.
We first learn an embedding model for each language,
 namely a high-resource language (English) and a low-resource one (Tagalog), 
independently.
Then, we learn a mapping between the two latent spaces to transfer the detection model.
We demonstrate that the proposed approach significantly outperforms state-of-the-art models, including BERT, and yields marked advantages in settings with very limited training data---the norm when dealing with detecting malicious activity in online platforms.
\end{abstract}





\maketitle

\input{src/introduction.tex}
\input{src/relatedworks.tex}

\input{src/data.tex}

\input{src/methodology.tex}
\input{src/results.tex}

\input{src/conclusion.tex}
\balance

\bibliographystyle{ACM-Reference-Format}
\bibliography{references}

\end{document}

%% file: src/introduction.tex
\section{Introduction}
\label{sec:intro}

Disinformation and political manipulation have a long history: for example, in 1984, long before the social media era, a story claiming that the HIV virus was created by the US government as a biological weapon became viral worldwide. %---this possibly represents the first popular instance of \textit{fake news}.
Nowadays, social media amplify and accelerate information spread   to an unprecedented pace. 
Online Social Networks (OSNs) like Twitter and Facebook have been facing a copious growth of malicious content, which undermine the truthfulness and authenticity of online discourse \cite{ferrara2019history,allcott2017social,grinberg2019fake,ferrara2015manipulation,vosoughi2018spread,woolley2017computational}.

Various studies showed that OSNs have been used for malicious purposes harming several constituents of our society \cite{luceri2021social,wang2022identifying}, ranging from geo-political events \cite{ferrara2017disinformation,luceri2019evolution,ferrara2020characterizing, pierri2022propaganda, sharma2022characterizing} to public health \cite{nogara2022disinformation, ferrara2020misinformation,chen2022charting,yang2021covid,ferrara2020types}.
Bots and trolls act as main actors in social media manipulation and disinformation campaigns \cite{Badawy2018,shao2018spread,luceri2020detecting,chang2022comparative,ferrara2022twitter}, often in a coordinated fashion \cite{ferrara2016rise,suresh2023tracking,sharma2021identifying,nizzoli2021coordinated,pacheco_2020_unveiling,weber2021amplifying}.

% \begin{figure}[t!] 
% \centering
%   \includegraphics[width=\columnwidth]{images/twitter.pdf}
%   \caption{Twitter account challenges. There has been a steady rise in the number of accounts suspended on Twitter to respond to the platform's misuse by malicious actors.}
%   \label{fig:twitter}
% % \vspace{-.5cm}
% \end{figure}

Particular attention has been devoted to the risk of mass manipulation of public opinion in the context of politics, whose prime example is the online interference in the 2016 US Presidential discussion election~\cite{bessi2016social,badawy2018falls}.
%On the other hand, 
Since then, OSNs have been trying to fight abuse and maintain a trustful and healthy conversation on their platforms.
% \cite{twitter1} reported to identifying and challenging tens of millions of potentially spam or automated accounts between September 2017 and May 2018 \cite{twitter2018bans} (Fig. \ref{fig:twitter}).
Despite the effort, the activity of trolls and bots appears to persist \cite{im2019still,luceri2019red,twitter1}. For instance,
% Twitter focused on foreign information operations and their interference in political conversations.
Twitter identified and suspended malicious accounts originating from diverse countries, including Russia, Iran, Bangladesh, and Venezuela \cite{twitter2}, suggesting the presence of coordinated efforts to manipulate online discourse across countries and languages. Recently, \citet{pierri2022propaganda} documented evidence of platform
abuse and subsequent Twitter interventions \cite{pierri2022does} in the context of the ongoing conflict between Ukraine and Russia.
\rev{While others have explored the various strategies of malicious users in high-resource languages \cite{luceri2021down,luceri2019red,torres2022manipulating,torres2022manufacture} to enable their detection \cite{chavoshi2016debot,chavoshi2018model,gupta2014tweetcred,minnich2017botwalk}, here we present a novel approach using transfer learning to empower the automated identification of misbehaving accounts in low-resource languages.
}
\subsection*{Contributions of this work}

% In this paper, we examine the activity of malicious accounts in the context of low-resource languages. In particular, we discovered a significant amount of users sharing content in Tagalog and interfering into the Twitter political discussion  during the 2016 US Presidential election.
Our aim  is to investigate whether and to what extent textual content can be used as a proxy to detect malicious activity on social media, with a particular focus on accounts sharing messages in low-resource languages.
Overall, we aim at answering two main research questions:

\smallskip\noindent\textbf{RQ1:} \textit{Can we classify an account as malicious based only on their shared content?} We explore the effectiveness of learning word representations from tweets to identify suspended accounts. 
% leveraging classifiers trained on the learned embeddings.

\smallskip\noindent\textbf{RQ2:} \textit{Can we learn a model from a high-resource language (English) and transfer knowledge to a low-resource one (Tagalog) for detecting suspended accounts?} We investigate whether learning a mapping between two independently-trained word embeddings can be beneficial to identify misbehaving accounts.
% transfer knowledge from one \emph{source} language to a \emph{target} language.

% \noindent Our contributions can be summarized as follows:
% \begin{itemize}
%     \item We introduce a novel and challenging problem regarding malicious behavior in OSNs: we focus on detecting the activity that may affect the health of online conversations in low-resource languages.
    
%     \item We propose to detect malicious behavior by leveraging only the text shared by Twitter accounts. To this aim, we present two possible solutions based on monolingual and cross-lingual word embeddings, respectively, showing promising results and opening the way to further research on this theme.
% \end{itemize}

%% file: src/relatedworks.tex
\section{Related Work}

\subsection{Political Manipulation}

Social media has provided an online venue for users to connect and share ideas. However, social media networks like Twitter and Facebook can be used for malicious purposes \cite{ferrara2015manipulation}. Social media mass manipulation of public opinion is based on disinformation campaigns 
carried out by malicious actors, including social media bots and state-backed trolls \cite{bessi2016social,shu2017fake,howard2017junk,woolley2017computational,luceri2019red,luceri2020detecting,ezzeddine2022characterizing,monsted2017evidence,boichak2018automated,bovet2019influence,scheufele2019science}.

Despite the mitigation efforts, recent election-related research shows that the number of bots has not significantly reduced \cite{deb2019bots,luceri2019red} and indeed bots are becoming more sophisticated \cite{luceri2019evolution}. Thus, potentially malicious activity on social media can become an even more pervasive problem for political discourse. Especially for the spread of fake news, various studies showed how political leaning \cite{allcott2017social,luceri2019red},  polarization \cite{bail2018exposure,azzimonti2018social}, age \cite{grinberg2019fake}, and education \cite{scheufele2019science} can greatly affect fake news spread, alongside with other mechanisms that leverage emotions \cite{ferrara2015quantifying,ferrara2015measuring}, cognitive limits \cite{PENNYCOOK2018,pennycook2019cognitive}, and social network vulnerabilities \cite{ceylan2023sharing,tommasel2022recommender,ribeiro2020auditing}. Other work established that social media platforms were used as well to distort other elections \cite{ratkiewicz2011detecting,metaxas2012social,howard2016bots,ferrara2017disinformation,del2017mapping,stella2018bots} and other real-world events \cite{pierri2022propaganda,ferrara2018measuring,nogara2022disinformation}. 
% Especially for the spread of fake news, various studies showed how political leaning \cite{allcott2017social,luceri2019red}, age \cite{grinberg2019fake}, and education \cite{scheufele2019science} can greatly affect fake news spread, alongside with other mechanisms that leverage emotions \cite{ferrara2015quantifying,ferrara2015measuring} and cognitive limits \cite{PENNYCOOK2018,pennycook2019cognitive}. 

\subsection{Bots \& Trolls}
The article \textit{The Rise of Social Bots} \cite{ferrara2016rise} initially highlighted the issue of bots, or algorithmic automated accounts, on social media platforms. \cite{bessi2016social} focused on bot detection during the 2016 US Presidential election, finding that an estimated 400K accounts were likely automated and produced nearly 1 in 5 tweets in that political conversation. Since the 2016 US Presidential election, the US election system has been under scrutiny. 
Since then, social media networks have been trying to fight malicious actors to maintain an healthy conversation on their platform. Foreign actors were shown in \cite{dutt2018senator} to influence unsuspecting users on social media for the expressed purpose of sowing discord. 
Badawy \textit{et al.} \cite{Badawy2018} analyzed the Russian troll accounts on Twitter to understand their information warfare campaign, while Im \textit{et al.} \cite{im2019still} showed that the accounts are remaining. Indeed, \citet{zannettou2019let} described how the automated identification of human operators, as state-backed troll accounts, is a challenging, yet-unsolved task.

These and other studies that corroborated the issue of automated accounts and trolls being used for malicious purposes have been used to inform policy and new regulations. For example,  the \textit{Bot Disclosure and Accountability Act} of 2018\footnote{\url{https://www.congress.gov/bill/115th-congress/senate-bill/3127/text}} first directs the Federal Trade Commission to implement controls on the social media companies and secondly, it amends the Federal Election Campaign Act of 1971 to prohibit a campaign from impersonating human activity online.

%% file: src/data.tex
\section{Data: US 2016 Presidential Election}
\label{sec:data}
In this study, we use Twitter as a test-bed to detect the activity of malicious accounts focusing on the 2016 US presidential election.
The dataset, about 42 million tweets posted by almost 6 million distinct users, was first published by \cite{bessi2016social}.
%and the 2018 US midterm elections.
Tweets were collected through the Twitter Streaming API using %various keywords. 
%The dataset was collected using 
23 election keywords (5 for Donald Trump, 4 for Hillary Clinton, 3 for third-party candidates, and 11 for the general election terms). 
The collection was carried out between September 16, 2016 and October 21, 2016. 
From the set of collected tweets, duplicates were removed, which may have been captured by accidental redundant queries to the Twitter API. 
\rev{A list of the most popular keywords and associated number of tweets is given in Table \ref{tab:hashtags}. }
\rev{Although all the keywords are in English, tweets in other languages were collected.} 

% THIS IS TYPICALLY UNNECESSARY UNLESS REQUESTED BY JOURNAL RULES ETC.
% To the best of our knowledge, we have complied with all of the Twitter's terms of service and privacy policies\footnote{\url{https://developer.twitter.com/}}. Because we do not report out any identifying information of the users collected and we only show analysis on aggregated data no Institutional Review Board approval was required or obtained. Additionally, while we did do a bot detection analysis using Botometer, we wanted to include those accounts as well, so the accounts identified as bots were not removed. 

\begin{table}[t!]
\centering\small
\begin{tabular}{lc}
\hline
    \textbf{Hashtag} & \textbf{\# of Tweets}   \\ 
    
\hline

\#election2016  &   422,952  \\
\#VPdebate  &   1,031,972  \\
\#hillary  &   1,516,318  \\
\#trump  &   3,290,636  \\
\#neverhillary  &   1,063,545  \\
\#nevertrump  &   746,430  \\
\#garyjohnson  &   58,832  \\
\#jillstein  &   51,831  \\

\hline

\end{tabular} 
\vspace{5mm}
\caption{A representative subset of hashtags used in the data collection, along with the number of tweets emdedding the hashtag
% \vspace{-5mm}
\label{tab:hashtags}} 
\end{table}

We identified %the English posts from the non-English posts. 
over 60 different languages, with the highest number of tweets written in European languages.
In particular, there were over 37.6 million English tweets posted by nearly 5 million users. 
We found a noticeable number of tweets in Tagalog, an Austronesian language which is the first language of a quarter of the population of the Philippines, and the second language of more than half of the rest. As the fourth-most common language by number of speakers in the United States \cite{acs}, behind only English, Spanish, and Chinese, Tagalog represents the top low-resource language in our data by number of tweets. The US is also home to one of the largest population of Filipino emigrants living outside of the Philippines. Additionally, Tagalog's low-resource status is further confirmed by an analysis of the size of its Wikipedia---a common proxy for estimating the amount of digital resources in a language. Tagalog's Wikipedia is currently ranked 101st by number of articles,\footnote{https://meta.wikimedia.org/wiki/List\_of\_Wikipedias} in sharp contrast to its prevalence in our dataset. For this reason, we focus our attention on Tagalog as the target language in this work.

%% file: src/methodology.tex
\section{Methodology} \label{sec:background}
% We here describe the proposed methodological framework. We first discuss \textit{FastText} and \textit{MUSE}, two embedding algorithms (\S\ref{sub:embeddings}), then describe how we adapt them to our learning tasks (\S\ref{sub:tasks}).

\subsection{Word Representations}

\label{sub:embeddings}

% \subsubsection{FastText}
% \paragraph{{FastText.}}
To learn word embeddings and train classification models, we use the  FastText\footnote{\url{https://github.com/facebookresearch/fastText}} framework.
Instead of treating words as atomic units of text, FastText  represents words as a bag of \emph{character n-grams} \cite{bojanowski2017enriching}, wherein each n-gram has its own vector representation and a word is represented as the sum of its constituent character n-grams. This allows the model to adapt to morphologically rich languages with large vocabularies as well as generalize better from smaller training corpora.

Although neural network-based models have achieved considerable success at text classification tasks, they remain quite expensive to train and deploy. FastText utilizes a hierarchical softmax to serve as a fast approximation of the softmax classifier to compute the probability distribution over the given classes \cite{joulin2016bag}. Using feature pruning, quantization, hashing, and retraining to substantially reduce model size without sacrificing accuracy or speed, this approach allows the training of models on large corpora of text much faster than neural network-based methods \cite{joulin2016fasttext}.
%The key feature of FastText is its ability to produce a continuous vector representation for any words in a text. Once the text is projected into the embedded space, the learned representation can be used on any downstream NLP task, such as text classification.

%Despite neural network models have become extremely accurate in NLP tasks and, more specifically, in text classification, their training phase is quite expensive both in time and resource consumption.
%On the other hand, FastText proposes a solution that addresses the trade-off between classification performance and computational time, while offering a small-size model, which in turn is desirable for various applications, and remaining competitive with deep learning models.

% FastText is based on n-gram features with the option of using a hierarchical softmax to serve as a fast approximation of the softmax classifier that is used to compute the probability distribution over the given classes.
% The n-gram features are employed to learn word representations while taking into account morphology.
% In particular, each character n-gram is associated with a vector representation, and, in turn, a word is represented as a sum (\textit{bag}) of its character n-gram. 
% Using feature pruning, quantization, hashing, and retraining to substantially reduce model size without sacrificing accuracy or speed,, this approach allows to train models on large corpus of data and, most importantly, to represent words that were not found in the training corpus.

% \subsubsection{MUSE}

\subsection{Transfer Learning}
% \paragraph{MUSE}
Traditional machine learning approaches for natural language processing focus on training specialized models for specific tasks. However, this requires significant amounts of data which is hard to acquire for low-resource languages. This has historically elicited more research on high-resource languages (primarily European), which leads to more resources created for these languages, thus feeding the cycle. Transfer learning has recently arisen as a way to leverage knowledge learned from a source language (or source task) and utilize it to improve performance on a target language (or target task).

To address the scarcity of data in the target language under analysis in this work, we use MUSE\footnote{\url{https://github.com/facebookresearch/MUSE}}, a framework for aligning monolingual word embeddings from different languages in the same space and allowing transfer of knowledge between them. MUSE learns a mapping from the source to target space using Procrustes alignment to minimize the distance between similar words in the two languages \cite{lample2017unsupervised}. It accepts as input two sets of pretrained monolingual word embeddings (such as those learned by FastText), one for each language, and can learn a mapping between them in either a supervised or unsupervised fashion. The supervised method requires the use of a bilingual dictionary to assist in aligning the two embeddings together by identifying similar word pairs that should be close together in the shared space. In the absence of such a dictionary, the unsupervised alternative utilizes adverserial training to initialize a linear mapping between a source and a target space and to produce a synthetic parallel dictionary. \cite{conneau2017word} showed that this approach can be used to perform unsupervised word translation without the use of any parallel data, with results that in some cases outperform even prior supervised methods.

% It then uses the dictionary (or adverserial training
% , in case the dictionary is not provided) to map the embeddings from both languages into the same space. MUSE comes with a large number of pretrained word embeddings, both monolingual and multilingual, as well as a set of bilingual dictionaries to perform mapping between other languages as well.
% 

%The rationale behind MUSE  is to build a bilingual dictionary between two languages without using any cross-lingual supervision, by leveraging only embedding spaces of each language.
%We employ this method as it is particularly suited for low-resource languages, for which there only exists a very limited amount of parallel data. 

%To this aim, we use MUSE, a Python library that provides multilingual word embeddings (FastText embeddings aligned in a common space) and a large-scale high-quality bilingual dictionaries for training and evaluation.

% We use the fastText embeddings trained on Wikipedia, and create a dictionary based on an online lexicon.

% It is then possible to apply the same techniques proposed for supervised techniques, namely a Procrustean optimization. 
%The MUSE project offers also to obtain cross-lingual word embeddings in a
%supervised way using a bilingual dictionary or identical character strings.

\subsection{Learning Tasks}
\label{sub:tasks}

% Next, we describe two different approaches to the text classification and transfer learning tasks.
% We model this task as a binary text classification between the labels ``\textit{Suspended}'' and ``\textit{Not suspended}''. Each instance in this case is the set of tweets for a particular user account, with the accompanying label specifying the user's account status. 
% To perform this classification, we  try two different approaches.

% \subsubsection{Monolingual text classification}\label{sub:mono}
\paragraph{Monolingual text classification.}
In the first approach, we train independent text classification models for each language from scratch using their respective datasets. For  classification purposes, we use the FastText framework, which represents text as a bag of words (BoW) and averages their individual representations into a combined text representation. This text representation is then used as input to a linear classifier with a softmax function that computes the probability distribution over the label classes in order to make predictions.
%We use this as our baseline.

% \subsubsection{Transfer learning}\label{sub:transfer}
\paragraph{Transfer learning.}
In the second approach, we use transfer learning from the high-resource language with more data (English) to improve text classification accuracy on the low-resource language with fewer data (Tagalog). We first train unsupervised monolingual word embeddings for each language using FastText's skipgram model \cite{mikolov2013efficient}. We then obtain multilingual word embeddings by mapping the embeddings for both English and Tagalog in the same space with MUSE by using a bilingual English-Tagalog dictionary to establish correspondence between words in the two languages, and eventually using these words as anchors to align the embeddings of both languages in the same latent space. This allows to maximize information learned from one language to another. The multilingual embeddings are then used as pretrained vectors to initialize a FastText model, trained on the target language using its dataset to make predictions over users' account status.

\subsection{Baseline Models}

We compare our work with a number of different baselines, both traditional and deep learning-based approaches, which we detail as follows.

\paragraph{Bag-of-Words and their TFIDF}
We create a bag-of-words model by extracting a vocabulary of words in the corpus. We then calculate the counts of the words in the examples as features for our model. For the TFIDF (term frequency---inverse document frequency) variant, we normalize the aforementioned counts  by dividing them by the total number of words in a document, which gives us the term-frequency. The inverse document frequency measures how common or rare a word is, and is equivalent to the logarithm of the total number of documents divided by the number of documents containing that word. The TFIDF is then given by the product of the term frequency and the inverse document frequency.

\paragraph{Bag-of-ngrams and their TFIDF}
% Often it is more informative to look at phrases of text instead of just single words in order to get more context. 
Often, sequences of words ($n$-grams) carry more information than words taken individually, specifically because n-grams carry contextual information that is lost when single words are considered.
We construct a bag-of-ngrams model by extracting n-grams, ranging from unigrams to 5-grams, ($n=1\dots 5$) from the corpus, and use those as features. For the TFIDF variant, we apply the same normalizing scheme as described above.

\paragraph{BERT contextual embeddings.}
Traditional word embeddings, while very efficient to compute and use, lack of context. As words exhibit polysemy, they can have different meanings based on the context in which they are used. An example is the word `bank', which can either mean a financial institution or the land alongside a river or lake. Recently, there has been a significant interest in the use of contextual word embedding models such as ELMo (Embeddings from Language Models) \cite{peters2018deep} and BERT (Bidirectional Encoder Representations from Transformers) \cite{devlin2018bert}, which are trained by a class of deep neural network models called \emph{transformers}. The final layers of these models have been shown to effectively capture a high degree of semantic knowledge from the input text, which can subsequently be used for downstream or auxiliary tasks. For our work, we use the contextualized word representations produced by the multilingual variant of the BERT model, which has been pre-trained on 104 languages using their respective \textit{Wikipedia}s as corpora. We extract the features generated by the final layer of BERT and train a softmax layer on top for our binary classification task.

%% file: src/results.tex
\section{Experimental Setup}
\label{sec:experimental_setup}

\subsection{Design}

For the purpose of this study, we define a malicious account as a user account that has been suspended from the Twitter platform. Although this is an approximation, Twitter systematically reviews suspicious accounts, either in an algorithmic fashion or based on reports of inappropriate behaviors, such as spam, abuse, etc. Given the negative cost associated with suspending a possibly legitimate account, Twitter's suspensions are typically associated with serious, repeated instances of misbehavior. In the context of political discussions, for example, reasons for suspension include the use of automated accounts, or orchestrated attempts to bolster the visibility or support of a political candidate.

 We assign a label to each user as either ``Suspended'' or ``Not suspended'' based on querying the Twitter Search API, whose response can be: (i) 
\textit{account\_suspended}, if Twitter has removed the account; (ii)  \textit{not\_found}, if the account has been deleted by the owner;  (iii)   \textit{protected}, if the owner has been made it private (i.e., invisible to the public); or, otherwise (iv) \textit{active} (i.e., not suspended). We hence use class (i) \textit{account\_suspended} as the positive label (i.e., ``Suspended'') and the other classes \textit{active}, \textit{not\_found} and \textit{protected} as the negative label (i.e., ``Not suspended'').

 %isotivlabel (i.e., ``Suspended''), ane classd(iv) (\textit{active}) as negative label (i.e., ``Not Susp ndede We did this in February 2019, more than two years since the data was collected.

\begin{table}[t!]
\centering\small
%\vspace{-.3cm}
\begin{tabular}{lcc}
\hline
    \textbf{Description} & \textbf{English} & \textbf{Tagalog}     \\ 
    
\hline

\# of accounts	&	4,872,565   &   23,979		\\
\% of suspended    &   31\%    &31\% \\
\hline

\# of tweets	&	37,623,535  &   29,887		\\
\% from suspended     &   29\%    & 31\% \\
\hline

\end{tabular} 
\vspace{5mm}
\caption{Statistics of the monolingual datasets. The amount of data for Tagalog is an order of magnitude less than that for English, but both languages contain a significant fraction of accounts that got suspended in the wake of the campaign. Furthermore, Tagalog data is also very sparse as we only have around 1 tweet per account, which also contributes to the difficulty of our classification task.
\label{tab:data}} 
% \vspace{-5mm}
\end{table}

\subsection{Tweet Aggregation}

We then aggregate all the tweets written in either English or Tagalog and build two monolingual datasets accordingly.
%\rev{such that the topic remained the same}.
In Table \ref{tab:data}, we list some statistics about users and tweets in the two sets of data.
For each dataset, we then aggregate tweets by user account by concatenating all text that a particular account has tweeted. This results in a tweet ``document'' for each user account. Such a set of tweets is then collectively used to make the prediction of whether a user was suspended or not. We minimize pre-processing the text in order to preserve characteristics of the tweet, such as punctuation, URLs, and hashtags, which can help flag potentially malicious users.

\subsection{Hyperparameters}

%We assign a label of either \textit{Suspended} or \textit{Not Suspended} to both English and Tagalog tweet authors. 
For both languages, we randomly sample 80\% of the data for training and retain the remainder for testing.

For the bag-of-words and n-gram-based models, we use approximately 35,000 features that are obtained from the text and train a logistic regression classifier with L2 regularization for 100 epochs on top of them.

For the BERT-based model, we use the the 768-dimensional contextual word embeddings obtained from the final layer of the model. We average the embedding for each token to produce a representation of the entire sequence. We feed this representation to a softmax layer with binary outputs, which we train for 100 epochs using the Adam optimizer \cite{kingma2014adam} with a learning rate of 0.001.

For the monolingual learning task, we train separate supervised FastText classification models for each language.
For the transfer learning task, we train separate unsupervised FastText word embeddings of dimensionality 100 using the skipgram model. We then align the two sets of word embeddings into the same space with MUSE and perform 5 iterations of refinement for alignment. The procedure uses the provided English-Tagalog dictionary which contains 5000 word pairs for training and 1500 for evaluation. We then use the aligned embeddings as pretrained vectors to initialize another FastText model for the target language (Tagalog) and compare it with a classic monolingual model. To evaluate the effectiveness of transfer learning in a low-resource setting, we train both models on only 10\% of the original Tagalog training set.

\subsection{Metrics}

We use F1, Precision, and Recall with binary averaging to evaluate the performance of our models. While \textit{macro} and \textit{micro averaging} give high scores, they are calculated globally and disregard class imbalance (the majority of unsuspended accounts lead to inflated results). We are, however, focused on the task of accurately identifying malicious users (the positive label in our setup), and thus we use binary averaging which reports results for the positive class of suspended accounts.

%%%% TO BE INCLUDED:
%we then aggregate user's tweet into a document per user. 
%Minimal pre-processing was done on the corpus. 
%Using the FastText embedding, previously mentioned we performed a 100-dimensional embedding. We then trained a classifier model on those embeddings for an English only and a Tagalog only. The models were trained using a learning rate of XX and the number of epochs were XX. 

%%%%%%%%%%%%%%%%
\section{Results}
\label{sec:results}

In this section, we present the results related to both the monolingual and cross-lingual approaches. 

Let us first consider the monolingual models for both English and Tagalog languages.
Table \ref{tab:perf} shows the performance in terms of precision, recall, and F1 score.
Note that these evaluation metrics have been computed by considering a binary classification scenario, where the positive label is represented by the class ``Suspended''.
What stands out is that precision is higher than recall in both languages,
suggesting that the models learn a conservative classification schema that minimizes the costly false positives.
% correctly predict an account to be suspended if it has a high probability of actually being suspended. 
This yields models missing a large number of suspended accounts in both languages, which is most apparent in the case of English, where the precision is high ($>$70\%), but the recall is quite low ($<$20\%). For Tagalog, the figures are somewhat closer to each other, resulting in a slightly higher F1 score w.r.t. English.
Despite the low recall scores may appear as indicative of large margins of improvement, which is addressed by our transfer learning model, it is also worth noting that in a practical setting, a conservative model shall be preferred over a more aggressive detection system that may yield a high false positive rate.

\begin{table}[t!]
\centering\small
%\vspace{-.3cm}
\begin{tabular}{lccc}
\hline
\textbf{Language}   	&	\textbf{Precision}	&	\textbf{Recall}	&	\textbf{F1}		\\
    
\hline
English	&	0.708	&	0.184	&	0.292		\\
% Tagalog	&	0.450	&	0.237	&	0.311		\\
Tagalog	&	0.448	&	0.218	&	0.293  \\
%Transfer	&	0.41	&	0.30	&	0.35	&	0.56	\\

\hline
\end{tabular} 
\vspace{5mm}
\caption{Performance of Monolingual Models. While F1 scores for both models are similar, the Tagalog model suffers in terms of precision in identifying suspended accounts.
\label{tab:perf}} 
% \vspace{-5mm}
\end{table}

%Two facts are worth noting: (i) precision is higher than recall in both the languages, and (ii) the gap between precision and recall is less noticeable in the Tagalog scenario, which in turn achieves a higher F1 score.
% indicating that the model better recognizes negative instances (not suspended) than positive

%In Table \ref{tab:transfer},  we show the performance of transfer learning (\S\ref{sub:transfer}) compared to the monolingual model when trained on a smaller (10\%) training set. Here, we see that the F1 score for the transfer learning model is much higher (more than twice) than that for the model trained from scratch, showing that transferring knowledge learned from one language to another within the same setting can help improving the classification accuracy.

\begin{figure}[t!] 
\centering
  \includegraphics[width=\columnwidth]{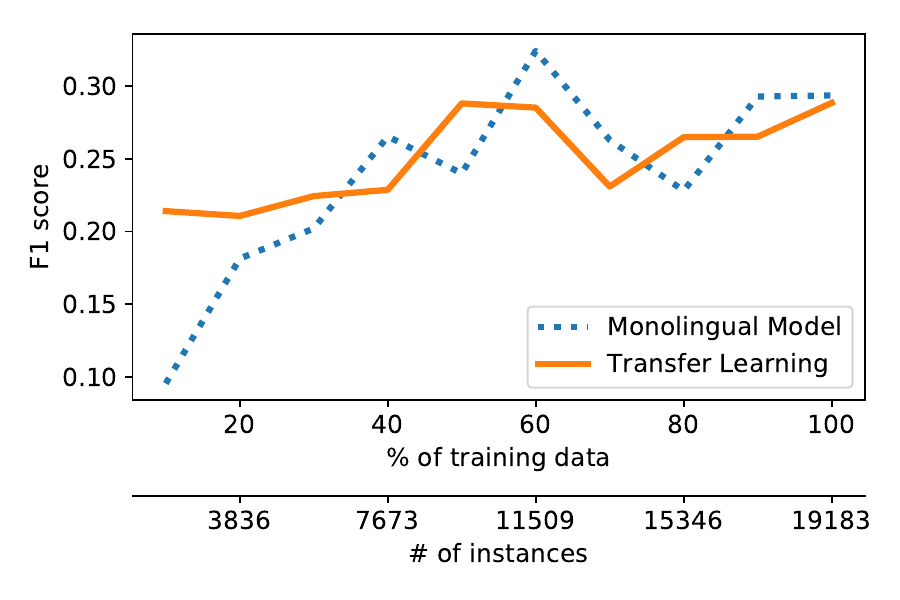}
% \vspace{-1cm}
  \caption{Monolingual vs. transfer learning at varying training set size. While enough data allows both to converge to similar performance, in low-resource cases, transfer learning vastly outperforms the monolingual approach.}
  \label{fig:transf_vs_mono}
% {-5mm}
\end{figure}

In Figure \ref{fig:transf_vs_mono}, we show the performance of transfer learning compared to the monolingual model at a varying training set size. 
%Performance is measured in terms of F1 score.
% as measured by the F1 score compares to that of the monolingual model with increasing training set size. 
Two facts are worth noting.
First, as we expected, as the percentage of training data increases, the F1 score of both models improves. Second, in cases where up to 30\% of the training set (5,754 instances) is used to train the models, transfer learning \rev{performs comparable to} %outperforms
the monolingual model and achieves a higher F1.

To further investigate this finding, in Table \ref{tab:transfer}, we show precision and recall of both the models in the low-resource setting where we use only 10\% of the Tagalog data as the training set. The monolingual model achieves better precision than transfer learning, but it performs poorly in terms of recall. However, the transfer learning model offers a more balanced trade-off between precision and recall, and appears to be more suitable in a low-resource setting.

We also evaluate the performance of our approach by comparing it with baseline models, as displayed in Table \ref{tab:modelcomparison}. To resemble the training setup for the baselines, here we train our proposed models for 100 epochs with a learning rate of 1.0. We see that the transfer learning approach outperforms all other methods with a higher F1 score and recall. While the baselines have good precision, they are unable to accurately detect malicious users, which is evident from their low recall. Our proposed approach shows promising performance especially when compared to a sophisticated model that uses contextualized word representations (BERT), probably because of the different nature of social media discourse with respect to the corpora BERT has been pre-trained on. This shows how the analysis of social media content, in general, and the detection of discourse manipulation, in particular, can not be easily conducted by using models trained on other, more formal bodies of text, further highlighting the downstream challenges of designing language-agnostic tools that can generalize to low-resource languages.

\rev{Finally, we perform dimensionality reduction of text embeddings using \textit{Uniform Manifold Approximation and Projection} (UMAP) \cite{mcinnes2018umap}. % and t-SNE \cite{maaten2008visualizing}. 
After aggregating the tweets for each user, we use FastText to generate vector embeddings from their text. We then use UMAP to map these embeddings into a lower-dimensional space, retaining only the components that explain most of the variation in the data. %, and label the users with their account status.
 Figure 2 shows the results of this projection onto a two-dimensional space. A separation is seen between the embeddings of the suspended and non-suspended tweets.} This suggests that the model can, to some degree, capture the distinction between the two classes of accounts in the target low-resource language.

\begin{table}[t!]
\centering\small
%\vspace{-.3cm}
\begin{tabular}{lccc}
\hline
\textbf{Model}   	&	\textbf{Precision}	&	\textbf{Recall}	&	\textbf{F1}		\\
    
\hline
FastText	&	0.847	&	0.051	&	0.095		\\
FastText + MUSE	&	0.390	&	0.147	&	0.214		\\
%Transfer	&	0.41	&	0.30	&	0.35	&	0.56	\\

\hline
\end{tabular} 
\vspace{5mm}
\caption{Models comparison in the low-resource setting. Evaluated on 10\% of the Tagalog training set, the transfer learning approach achieves more than twice the F1 score and almost three times the recall of the more conservative monolingual approach.
\label{tab:transfer}} 
% \vspace{-5mm}
\end{table}

\begin{table}[t!]
\centering\small
%\vspace{-.3cm}
\begin{tabular}{lccc}
\hline
\textbf{Model}   	&	\textbf{Precision}	&	\textbf{Recall}	&	\textbf{F1}		\\
    
\hline
Bag-of-Words	&	0.502	&	0.172	&	0.256		\\
Bag-of-Words + TFIDF	&	0.580	&	0.125	&	0.206		\\
N-grams &	0.546	&	0.166	&	0.247		\\
N-grams + TFIDF	&	\textbf{0.635}	&	0.104	&	0.179		\\
BERT embeddings	&	0.513	&	0.136	&	0.215		\\
\textbf{FastText}	&	0.424	&	0.237	&	0.337		\\ % 50 epochs
\textbf{FastText + MUSE}	&	0.416	&	\textbf{0.280}	&	\textbf{0.347}		\\ % 50 epochs
% FastText	&	0.448	&	0.218	&	\textbf{0.293}		\\
% FastText + MUSE	&	0.420	& \textbf{0.219}	&	0.288		\\
%Transfer	&	0.41	&	0.30	&	0.35	&	0.56	\\

\hline
\end{tabular} 
\vspace{5mm}
\caption{Model comparison on Tagalog. Among our proposed methods (in bold), transfer learning outperforms all other models in terms of F1 and recall.
\label{tab:modelcomparison}} 
% {-5mm}
\end{table}

\begin{figure}[t!] 
\label{fig:umap}
\centering
  \includegraphics[width=\columnwidth, clip, trim=25 15 25 25]{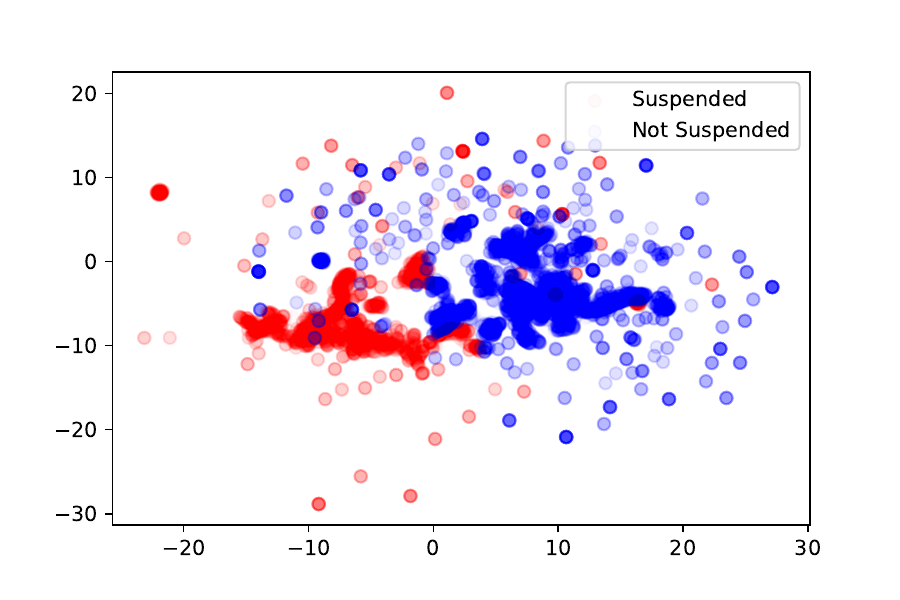}
% \vspace{-5mm}
%   \caption{Visualization of tweets with UMAP}
% \vspace{-5mm}
% }
\caption{Visualization of Tweets. UMAP shows a clear separation between the text embeddings of suspended (red) and not suspended (blue) accounts.}
\end{figure}

% \begin{table}[t!]
% \begin{center}
% \begin{tabular}{|l|rl|}
% \hline \textbf{Type of Text} & \textbf{Font Size} & \textbf{Style} \\ \hline
% paper title & 15 pt & bold \\
% author names & 12 pt & bold \\
% author affiliation & 12 pt & \\
% the word ``Abstract'' & 12 pt & bold \\
% section titles & 12 pt & bold \\
% subsection titles & 11 pt & bold \\
% document text & 11 pt  &\\
% captions & 10 pt & \\
% abstract text & 10 pt & \\
% bibliography & 10 pt & \\
% footnotes & 9 pt & \\
% \hline
% \end{tabular}
% \end{center}
% \caption{\label{font-table} Font guide. }
% \end{table}

%% file: src/conclusion.tex
\section{Conclusions}
\label{sec:conclusion}

Since billions of users populate social media platforms around the world, these represents ripe targets for malicious actors and deceptive behaviors. We can leverage NLP to assist in an automated way
%to reduce some of the burden of detecting 
the detection of manipulation efforts.
Importantly, the research community has predominantly been focusing on the study of online platforms in high-resource languages, for which many NLP tools exist, e.g., sentiment analysis, semantic parsers, etc. 
% that facilitate the work. 
However, a need for language-agnostic frameworks exists to allow the study of discourse in low-resource languages and enable the automated identification of malicious activity.
In this paper, we posed the problem of detecting social media abuse in low-resource languages. 

Using a backdrop of the 2016 US Presidential election Twitter discussion, and by drawing a parallel between abuse in English and Tagalog, we proposed a framework to detect suspended, misbehaving accounts leveraging only their shared content in Tagalog. Although the task was proven to be challenging even in a high-resource language setting, we showed that our proposed framework can build a conservative model to detect malicious actors manipulating the discourse in a low-resource language. 
Much more work is needed to guarantee healthy conversations in the online landscape of low-resource languages. New language-agnostic tools can spur from our work. We seek to initiate an agenda and stimulate research on social media in low-resource languages, including for the study of misbehavior, manipulation, and abuse.

% posed the problem of detecting social media abuse in low-resource languages